\documentclass[graybox]{svmult/styles/svmult}

\usepackage{acronym} 
\usepackage{mathptmx}       
\usepackage{helvet}         
\usepackage{courier}        
\usepackage{type1cm}        
\usepackage{makeidx}         
\usepackage{graphicx}        
\usepackage{multicol}        
\usepackage[bottom]{footmisc}
\usepackage[acronym,xindy,toc,nonumberlist,shortcuts]{glossaries}

\newacronym{vgi}{VGI}{Volunteered Geographic Information}
\newacronym{osm}{OSM}{OpenStreetMap}
\newacronym{hci}{HCI}{human-computer interaction}
\newacronym{gir}{GIR}{Geographic Information Retrieval}
\newacronym{ai}{AI}{artificial intelligence}
\newacronym{ir}{IR}{information retrieval}
\newacronym{gis}{GIS}{Geographic Information Systems}
\newacronym{idf}{IDF}{Inverse Document Frequency}
\newacronym{lsa}{LSA}{Latent Semantic Analysis}
\newacronym{poipl}{POI}{points of interest}
\newacronym{poi}{POI}{point of interest}
\newacronym{vsm}{VSM}{Vector Space Model}
\newacronym{pos}{POS}{part-of-speech}
\newacronym{skos}{SKOS}{Simple Knowledge Organization System}
\newacronym{rdf}{RDF}{Resource Description Framework}
\newacronym{owl}{OWL}{Web Ontology Language}
\newacronym{giscience}{GIScience}{Geographic Information Science}
\newacronym{wsddef}{WSD}{Word Sense Disambiguation}
\newacronym{dl}{DL}{description logic}
\newacronym{mdsmsim}{MDSM}{Matching-Distance Similarity Measure}
\newacronym{oursurvey}{GeReSiD}{Geo Relatedness and Similarity Dataset}
\newacronym{ira}{IRA}{interrater agreement}
\newacronym{bow}{BOW}{bag-of-words}
\newacronym{irr}{IRR}{interrater reliability}
\newacronym{mds}{MDS}{multidimensional scaling}
\newacronym{lbs}{LBS}{location-based services}
\newacronym{oss}{FOSS}{free and open-source software}
\newacronym{sdts}{SDTS}{Spatial Data Transfer Standard}
\newacronym{tfidf}{TF-IDF}{Term Frequency-Inverse Document Frequency}
\newacronym{ogc}{OGC}{Open Geospatial Consortium}

\newglossaryentry{w2}{name=Web 2.0,
	description={TODO}
}

\newglossaryentry{stratag}{name=Strategic Research in Advanced Geotechnologies (StratAG)}

\newglossaryentry{kb}{name={knowledge base}}
\newglossaryentry{gkb}{name={geo-knowledge base}}

\newglossaryentry{nuim}{name={National University of Ireland, Maynooth},
	description={TODO}
}

\newglossaryentry{ucd}{name={University College Dublin},
	description={TODO}
}

\newglossaryentry{osmsim}{name=\textsc{OSM-TagSim},
	description={TODO}
}

\newglossaryentry{lexsim}{name=\textsc{osm-sim_{lex}},
	description={TODO}
}

\newglossaryentry{simdl}{name=Sim-DL,
	description={TODO}
}

\newglossaryentry{netsim}{name=\textsc{osm-sim_{sim}},
	description={TODO}
}

\newglossaryentry{dbp}{name=DBpedia,
	description={TODO}
}

\newglossaryentry{lgd}{name=LinkedGeoData,
	description={TODO}
}

\newglossaryentry{sgw}{name=Semantic Geospatial Web,
	description={TODO}
}

\newglossaryentry{sw}{name=Semantic Web,
	description={TODO}
}

\newglossaryentry{webplatform}{name=Web platform for map personalisation and visualisation,
	description={TODO}
}

\newglossaryentry{wsd}{name=word sense disambiguation,
	description={TODO}
}

\newglossaryentry{os}{name=open source,
	description={TODO}
}

\newglossaryentry{osn}{name=OSM Semantic Network,
	description={TODO}
}

\newglossaryentry{nlp}{name=natural language processing,
	description={TODO}
}

\newglossaryentry{owc}{name=OSM Wiki Crawler,
	description={TODO}
}

\newglossaryentry{oww}{name=OSM Wiki website,
	description={\url{http://wiki.openstreetmap.org}}
}

\newglossaryentry{mdsm}{name=MDSM evaluation dataset,
	description={TODO}
}

\newcommand{\homepage}[1]{\url{http://github.com/ucd-spatial/#1} (acc. June 5, 2012)}
\newcommand{\footurl}[1]{\footnote{\url{#1} (acc. June 5, 2012)}}
\newcommand{\graphdate}[0]{June 10, 2011}

\usepackage{array}
\newcolumntype{P}[1]{>{\raggedright\arraybackslash}p{#1}}

\makeindex             

\begin{document} 

\title*{A Survey of Volunteered Open Geo-Knowledge Bases in the Semantic Web}
\author{Andrea Ballatore, David C. Wilson and Michela Bertolotto}
\institute{Andrea Ballatore \at School of Computer Science and Informatics, University College Dublin, Belfield, Dublin 4, Ireland, \email{andrea.ballatore@ucd.ie}
\and David C. Wilson \at Department of Software and Information Systems, University of North Carolina, 9201 University City Boulevard, Charlotte, NC 28223-0001, USA, \email{davils@uncc.edu}
\and Michela Bertolotto \at School of Computer Science and Informatics, University College Dublin, Belfield, Dublin 4, Ireland, \email{michela.bertolotto@ucd.ie}}
%
%
\maketitle
\acresetall


\abstract{
Over the past decade, rapid advances in web technologies, coupled with innovative models of spatial data collection and consumption, have generated a robust growth in geo-referenced information, resulting in spatial information overload.
Increasing `geographic intelligence' in traditional text-based information retrieval has become a prominent approach to respond to this issue and to fulfill users' spatial information needs.
Numerous efforts in the \gls{sgw}, \gls{vgi}, and the Linking Open Data initiative have converged in a constellation of open \glspl{kb}, freely available online.
In this article, we survey these open \glspl{kb}, focusing on their geospatial dimension.
Particular attention is devoted to the crucial issue of the quality of \glspl{gkb}, as well as of crowdsourced data.
A new \gls{kb}, the OpenStreetMap Semantic Network, is outlined as our contribution to this area.
Research directions in information integration and \gls{gir} are then reviewed, with a critical discussion of their current limitations and future prospects.
}

\section{Introduction}
\acresetall 
\label{sec:digearth}
In 1998, U.S. Vice President Al Gore delivered a speech at the California Science Center about what he named Digital Earth, a ``multi-resolution, three-dimensional representation of the planet, into which we can embed vast quantities of geo-referenced data'' \cite[p. 89]{gore:1998:digital}.
Much of the unprecedented amount of information produced and released on the Internet is about a specific place on the Earth.
However, Gore pointed out, most of this informational wealth is generated and left untapped.
Among the key aspects that would enable a more efficient exploitation of geo-data, interoperability and metadata were considered of particular importance.
Multiple data sources should be combined together using a common framework, and metadata should describe online resources in a clear, standardised way \cite{gore:1998:digital}.

Over the past 14 years, several geospatial initiatives have been undertaken, oriented towards the implementation of the Digital Earth \cite{goodchild:2009:neogeography}.
The \gls{ogc}\footurl{http://www.opengeospatial.org} has defined and promoted several standards to distribute geographic data, while the Global Spatial Data Infrastructure (GSDI) association aims at fostering ``spatial data infrastructures that support sustainable social, economic, and environmental systems integrated from local to global scales'' \cite[p. 1]{gsdi:2009:strategic}.
Despite these efforts, standard formats are often ignored in favour of application-specific formats.
As Fonseca et al. put it, heterogeneity emerges spontaneously in a free market of ideas and products, and standards cannot reduce it by decree \cite{fonseca:2002:using}.

The spectacular growth of unstructured information online has affected all domains, prompting Tim Berners-Lee to envisage the advent of the so-called Semantic Web \cite{Berners:2001:semantic}.
The Semantic Web project aims to develop a standard semantic format to describe online data, originating a network of machine-readable, semantically clear documents.
Data semantics is expressed in predicate logic-based languages such as RDF,\footurl{http://www.w3.org/RDF} in large collections of statements about real world entities.  
This vision was further formalised through the Linked Data initiative, which promotes the release of datasets in an inter-connected web of semantic data \cite{Bizer:2009:linked}.


The information explosion in geographic data has not only been quantitative, but also qualitative \cite{xiao:2009:intelligent}.
With the rise of Web 2.0, Internet users have become active producers of geo-referenced information, utilising collaborative web tools in large projects \cite{Oreilly:2005:web20}.
Several collaborative efforts emerged to create and maintain large datasets, resulting in crowdsourcing, impacting initially on non-spatial information and subsequently also on the geographic domain \cite{Howe:2006:rise}.
In the geospatial context, the term `neogeography' has been used in order to refer to this rapid and complex nexus of technological and social practices \cite{Turner:2006:introduction}.
\glsreset{vgi}
Goodchild termed the crowdsourcing of geographic information as \gls{vgi}, emphasising its production through voluntary labour \cite{Goodchild:2007:citizens}.
Haklay et al. have surveyed VGI projects \cite{Haklay:2008:web}, while Coleman et al. have discussed the practices and motivations of `produsers', users/producers of geographic data \cite{coleman:2009:volunteered}.
In addition, Sui used the term `wikification' to describe the practice of crowdsourcing of non-textual data, emulating the Wikipedia model in the geographic domain \cite{Sui:2008:wikification}.
 
The impact of neogeography is not restricted to non-profit, academic organisations.
Private institutions such as Google, Microsoft, and Yahoo! are progressively offering facilities for sharing geo-data, expanding their services beyond the routing systems that dominated the first phase of web-based \gls{gis}.\footnote{See \url{http://maps.google.com}, \url{http://www.bing.com/maps}, \url{http://maps.yahoo.com}}
In this sense, geo-wikification is identifiable in the growth of web services allowing users, with some degree of freedom, to create or edit spatial data.
As Priedhorsky notes, however, most interactive geo-services are essentially `digital graffiti,' i.e. annotations on a static geographic image \cite{priedhorsky:2008:computational}.
Beyond the specificities of each case, it can be argued that all neogeographic and VGI phenomena share the characteristics of being volunteered, crowdsourced, wikified, and web-based.


Even though the popular claim that 80\% of information is geo-referenced has been questioned \cite{hahmann:2011:all}, it can be stated safely that, over the past decade, geo-information has experienced a remarkable growth \cite{kitsuregawa:2010:special}.
As happened in other fields subject to an information explosion and subsequently to information overload, the issue of semantics of geo-data -- or lack thereof -- has become critical.
The deluge of semantically ambiguous geo-data caused Egenhofer to advocate the emergence of a \gls{sgw}, a spatial extension of the Semantic Web \cite{Egenhofer:2002:toward}.
In Egenhofer's view, this new framework for geospatial information retrieval should rely on the semantics of spatial and terminological ontologies.
Thanks to inter-operable semantic representations of the data, the \gls{sgw} will increase the relevance and quality of results in geographic retrieval systems.

As a result of the synergy between crowdsourcing, VGI, and the \gls{sgw}, several large-scale collaborative projects have emerged.
While Wikipedia\footurl{http://www.wikipedia.org} is without doubt the most visible text-based crowdsourcing project, \gls{osm} has applied the wiki model to create an open world vector map \cite{Haklay:2008:openstreetmap}.
Several \glspl{gkb} have then been created by structuring existing datasets into Semantic Web formats: the projects LinkedGeoData, GeoNames, and GeoWordNet are salient examples \cite{Auer:2009:linkedgeodata,giunchiglia:2010:geowordnet}.
Research efforts have been undertaken on the development, maintenance, and merging of open \glspl{gkb}, to enhance the geographic intelligence of information retrieval systems, beyond the traditional text-based techniques \cite{fonseca:2002:using,xiao:2009:intelligent,Auer:2009:linkedgeodata,euzenat:2010:results}. 

Moreover, \gls{gir} has attempted to increase the geographic awareness of text-based information retrieval systems.
On top of traditional flat gazetteers (dictionaries of toponyms and geo-coordinates), \gls{gir} has started exploiting \glspl{gkb} to reduce the ambiguity of geographic terms and enable spatial reasoning \cite{jones:2001:geographical,Overell:2009:geographic}.
Despite these efforts, the knowledge contained in such computational artifacts is left largely untapped.
We believe that these open \glspl{gkb} have potential in addressing the challenges of GIR, and deserve particular attention.
For this reason, we provide a survey of currently active \glspl{kb} with particular emphasis on their geospatial content, and we review the state of the art in information integration and \gls{gir}, including our contribution to these areas.



%
The remainder of this chapter is organised as follows: Section \ref{sec:survey} surveys the constellation of online open \glspl{kb} containing geographic knowledge.
Applications of \glspl{gkb} are discussed in Section \ref{sec:action}, from recent efforts in ontology alignment and merging (Section \ref{sec:aligning}), to ontology-powered \gls{gir} (Section \ref{sec:gir}).
Section \ref{sec:osmsemnetwork} presents our work in this area, describing the \gls{osn}\footurl{http://wiki.openstreetmap.org/wiki/OSMSemanticNetwork} and the semantic expansion of the \gls{osm} dataset, connecting it to DBpedia, in order to enrich spatial data with a richer \gls{kb} \cite{Ballatore:2011:semantically}.
The issue of quality of \glspl{gkb} is discussed in Section \ref{sec:quality}, and Section \ref{sec:limitations} offers a review of current limitations of these computational artifacts that need particular consideration in their usage.
Finally, Section \ref{sec:concl} discusses of the challenges lying ahead in this field and the further research required to identify solutions to these challenges.

\section{Survey of Open Linked Geo-Knowledge Bases}
\label{sec:survey}

This section provides a survey of open, collaborative \glspl{gkb}, which constitute an important part of semantic technologies.
To avoid terminological confusion, it is beneficial to provide a definition of the related and sometimes overlapping terms used in knowledge representation. 
A `knowledge base' is a collection of facts about a domain of interest, typically organised to perform automatic inferences \cite{guarino:1995:ontologies}.
A knowledge base contains a terminological conceptualisation (typically called `ontology') and a set of individuals.
Widely used both in philosophy and in computer science, the meaning of the term `ontology' is particularly difficult to define \cite{smith:2003:ontology}.
Among the many definitions, ``an explicit specification of a conceptualization'' and ``shared understanding of some domain of interest'' are of particular relevance, as they stress the presence of an explicit formalisation, and the general aim of being understood within a given domain \cite[p. 587]{guarino:1995:ontologies}.
Winter notes that ontologies became part of \gls{giscience} towards the end of the 20th century \cite{winter:2001:ontology}.

A `thesaurus' is a list of words grouped together according to similarity of their meaning \cite{roget:1998:roget}, whilst a digital `gazetteer' is specifically geographic, and contains toponyms, categories, and spatial footprints of geographic features \cite{hill:2000:core}.     
In the \gls{w2} jargon, a `folksonomy' is a crowdsourced classification of online objects, based on an open tagging process \cite{vander:2007:folksonomy}.
Finally, a `semantic network,' a term which originated in psychology, is a graph whose vertices represent concepts, and whose edges represent semantic relations between concepts \cite{rada:1989:development}.\footnote{Unlike ontologies, semantic networks focus on psycho-linguistic aspects of the terms.  However, some \glspl{kb}, such as WordNet, defy this distinction by showing aspects of both ontologies and semantic networks.}
We define a `\gls{gkb}' as a \gls{kb} containing some geographic information.

In the context of geographic information, a \gls{kb} is generally made up of an ontology, defining classes and their relationships (abstract geographic concepts such as `lake'), and then populated with instances of these classes, generally referring to individual entities (e.g. Lake Victoria and Lake Balaton).
In this survey, we restrict the scope to projects having global coverage, discussing their spatial content.
These \glspl{kb} are the result of combined efforts in crowdsourcing, VGI, and the \gls{sgw}, and offer useful resources for \gls{gir}, and other areas of geo-science.

The Semantic Web and the Linked Data initiatives promote the adoption of semantic formats, which can be used to add an open, machine readable semantic structure to online data \cite{Bizer:2009:linked,goodwin:2008:geographical}.
In this context, several collaborative projects have emerged, resulting in a growing number of freely available \glspl{gkb}.
Among these numerous resources, we focus on eleven datasets that have a global scope (as opposed to local projects), are mostly generated through crowdsourcing, released under Creative Commons/Open Database licences,\footurl{See http://creativecommons.org and http://opendatacommons.org} and which are available as fully downloadable dumps in popular Semantic Web formats such as OWL and RDF.
Some of the selected projects are focused specifically on geographic data (e.g. GeoNames and OpenStreetMap), while others are more general-purpose but contain valuable geographic knowledge (e.g. DBpedia and Freebase).
These \glspl{kb} provide open datasets, and are inter-connected with one another.
Our own contribution to this area of research, the \gls{osn}, is described in Section \ref{sec:osmsemnetwork}.
Relevant characteristics of each \gls{kb} are summarised in Table \ref{tab:resources}.


\begin{table}[htbp]
	\begin{tabular}{lP{3.5em}P{13em}P{7em}P{5em}}
	$~$\\
	$~$\\
	$~$\\
	$~$\\
	$~$\\
	$~$\\
	$~$\\
	\hline\noalign{\smallskip} 
	\emph{Project Name} & \emph{Year*}  & \emph{Type \& Content} & \emph{Data sources} & \emph{Formats}\\
	\hline\hline
	 
\textsc{ConceptNet} & 2000 & Ontology, semantic network; 1.6 million assertions,  700,000 natural language sentences & Wikipedia, WordNet, and others & JSON\\ \hline 
\textsc{DBpedia} & 2007 & Ontology, semantic network; 320 classes, 740K Wikipedia types, 3.6M entities, 1 billion triples & Wikipedia & OWL/RDF\\ \hline 
\textsc{Freebase} & 2007 & Ontology, knowledge base; 22M+ entities, 1M locations & Crowdsourced & Tab separated text\\ \hline
\textsc{GeoNames} & 2006 & Gazetteer; 650 classes, 10M+ toponyms & Gazetteers, Wikipedia, crowdsourced & OWL/RDF\\\hline 
\textsc{GeoWordNet} & 2010 & Semantic network, thesaurus, gazetteer; 330 classes, 3.6M entities & WordNet, {GeoNames}, MultiWordNet & RDF\\ \hline
\textsc{LinkedGeoData} & 2009  & Gazetteer; 1K classes, 380M geographic entities & OpenStreetMap & RDF\\\hline 
\textsc{OpenCyc} & 1984 & Ontology, semantic network; 50K classes, 300K facts & Expert-authored Cyc knowledge base & OWL/RDF\\ \hline
\textsc{OpenStreetMap} & 2004  & Vector map, gazetteer; User-defined tags, 1.2B nodes, 114M ways & Crowdsourced, free GIS datasets & XML\\\hline 
\textsc{Wikipedia} & 2001  & Semantic network, dictionary, thesaurus; Semi-structured (infoboxes), 3.9M articles in English & Crowdsourced & XML\\\hline 
\textsc{WordNet} & 1985  & Semantic network, dictionary, thesaurus; 117K synsets & Expert-authored knowledge base & OWL/RDF\\ \hline
\textsc{Yago} & 2006  & Ontology, semantic network; 10M+ entities, 460M facts & Wikipedia, GeoNames, WordNet & RDF\\
	\noalign{\smallskip}\hline\noalign{\smallskip}
	\end{tabular}
	\caption{A survey of open ontologies. All of these projects are currently active, release open data, have global scope, and are interconnected with other projects. *Beginning of the project.}
	\label{tab:resources}
\end{table}

\begin{description}
\item[\textsc{ConceptNet}] This semantic network is focused on natural language processing and understanding \cite{havasi:2007:conceptnet}.
ConceptNet is a large semantic network, whose nodes represent concepts in the form of words or short phrases of natural language.
The graph edges represent labelled relationships.
Each statement in ConceptNet has justifications pointing to it, explaining where it comes from and how reliable the information seems to be.
The ontology includes 1.6 million assertions gathered from Wikipedia, Wiktionary, WordNet, and the 700,000 sentences from the Open Mind Common Sense project \cite{singh:2002:open}.
Efforts to encode ConceptNet in RDF are being undertaken \cite{grassi:2011:towards}.
 
\item[\textsc{DBpedia}] One of the leading projects of the Semantic Web, DBpedia is a Semantic Web version of Wikipedia \cite{Auer:2007:dbpedia}.
The knowledge base currently contains 3.6 million entities, encoded in a billion RDF triples, including 526,000 places.
As DBpedia is strongly interconnected with other \glspl{kb} (e.g. WordNet W3C, GeoNames, LinkedGeoData), it is considered the central hub of Linked Data.

\item[\textsc{Freebase}] Designed as an open repository of structured data, Freebase allows web communities to build data-driven applications \cite{bollacker:2008:freebase}.
The \gls{kb} is structured around terms (classes), and unique entities (instances), where an entity can be a specific person, a place, or a thing, and is described by facts.
It currently contains 22 million entities, of which 1 million are locations.
As entities are described by facts corresponding to a directed graph, it can be easily converted into RDF.
 
\item[\textsc{GeoNames}] Combining multiple data sources, GeoNames aims at offering a large, volunteered gazetteer.\footurl{http://www.geonames.org}
The \gls{kb} contains over 10 million toponyms, structured in 650 classes.
GeoNames integrates geographical data such as names of places in various languages, elevation, and population.
The data is collected from traditional gazetteers such as National Geospatial-Intelligence Agency's (NGA) and the U.S. Geological Survey Geographic Names Information System (GNIS), and crowdsourced online.

\item[\textsc{GeoWordNet}] Geo\-WordNet is the result of the integration of WordNet, GeoNames and the Italian part of MultiWordNet \cite{giunchiglia:2010:geowordnet}.  
It is a hybrid project, combining a semantic network, a dictionary, a thesaurus, and a gazetteer. 
  It was developed in response to the limited WordNet coverage of geospatial information and lack of concept grounding with spatial coordinates. The \gls{kb} contains 3.6 million entities, 9.1 million relations between entities, 334 geographic concepts, and 13,000 (English and Italian) alternative entity names, for a total of 53 million RDF triples.

\item[\textsc{LinkedGeoData (LGD)}] Since OpenStreetMap has gathered a large collection of geographic data, LinkedGeoData is an effort to republish it in the Semantic Web context \cite{Auer:2009:linkedgeodata}.
The \gls{osm} vector dataset is expressed in RDF according to the Linked Open Data principles, resulting in a large spatial knowledge base.
The knowledge base currently contains 350 million nodes, 30 million ways (polygons and polylines in the \gls{osm} terminology), resulting in 2 billion RDF triples.
Some entities are linked with the corresponding ones in DBpedia.

\item[\textsc{OpenCyc}] This is the open source version of Cyc, a long running artificial intelligence project, aimed at providing a general knowledge base and common sense reasoning engine.\footurl{http://www.cyc.com/opencyc, http://sw.opencyc.org} 
Even though OpenCyc covers a limited number of geographic instances, it contains a rich representation of specialised geographic classes, such as \emph{salt lake} and \emph{monsoon forest}.
The OpenCyc classes are interlinked with DBpedia nodes and Wikipedia articles.

\item[\textsc{OpenStreetMap (OSM)}] The \gls{osm} project aims at constructing a world vector map \cite{Haklay:2008:openstreetmap}.
The leading VGI initiative, the dataset represents the entire planet, gathering data from existing datasets, GPS traces, and crowdsourced knowledge.
To date, the vector dataset contains 1.2 billion nodes (points), and 115 million ways (polygons and polylines).

\item[\textsc{Wikipedia}] A collaborative writing project, Wikipedia is a multilingual, universal encyclopedia, and has become the most visible crowdsourcing phenomenon.\footurl{http://www.wikipedia.org}
The English version currently contains 3.9 million articles, resulting in a 2 billion-word corpus.
Because of high connectivity between its articles, Wikipedia is sometimes used as a semantic network \cite{strube:2006:wikirelate}.
This vast repository of general knowledge has been used for different purposes, including semantic similarity and ontology extraction \cite{Volkel:2006:semantic,nakayama:2008:wikipedia}.
The project has also attracted interest in the area of GIScience \cite{odon:2010:geographical}.

\item[\textsc{WordNet}] Initially conceived as a lexical database for machine translation, WordNet has become a widely used resource in various branches of computer science, where it is used as a semantic network and as an ontology \cite{fellbaum:1998:wordnet}.
Currently it contains 117,000 `synsets', groups of synonyms corresponding to a concept, connected to other concepts through several semantic relations.
The dataset has been encoded and released in RDF, becoming a highly linked \gls{kb} in the web of Linked Open Data.\footurl{http://www.w3.org/TR/wordnet-rdf}
Even though the spatial content of WordNet is limited, the ontology holds a high quality, expert-authored conceptualisation of geographic concepts.     

\item[\textsc{Yago}] Yet Another Great Ontology (YAGO) is a large knowledge base extracted from Wikipedia and Wordnet \cite{Suchanek:2007:yago}.
Recently YAGO has been extended with data from GeoNames, with particular emphasis on the spatial and temporal dimensions \cite{hoffart:2011:yago2}.
The current version of the \gls{kb} contains 10 million entities, encoded in 460 million facts.
YAGO is inter-linked with DBpedia and Freebase.   

\end{description}

\begin{figure}[t]
  \includegraphics[width=39em]{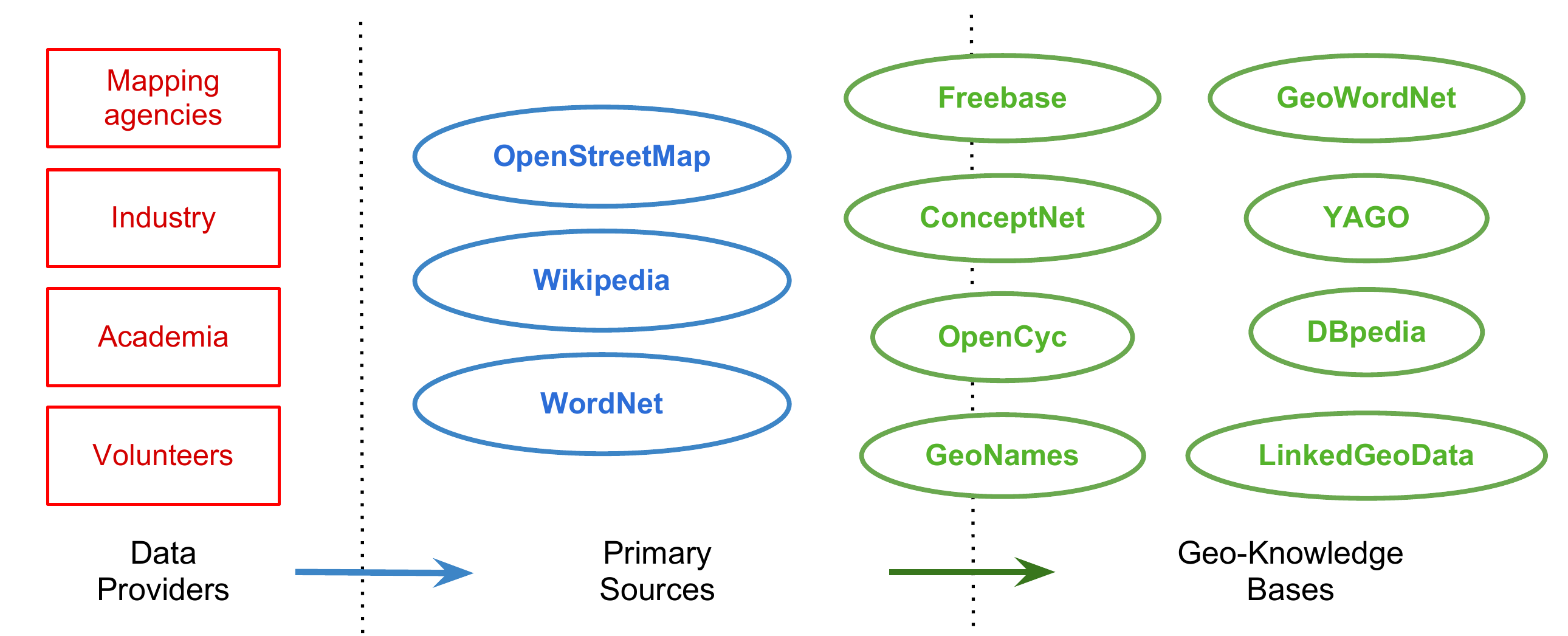}
 \caption[The constellation of open linked knowledge bases]{The constellation of open \glspl{gkb}. The data path is schematised from the data providers to semi-structured primary sources, and finally structured into \glspl{kb}. Some projects defy classification by producing new knowledge in structured \glspl{kb}, and extracting knowledge from primary sources.}
 \label{fig:constellation}       
 \end{figure}

Figure \ref{fig:constellation} presents the constellation of \glspl{gkb}, showing a schematised data path from the data producers to the \glspl{kb}.
Bearing in mind the complexity of these collaborative processes, the main actors in this constellation, involved in the production of information and the generation of open linked \glspl{kb}, can be grouped as follows:

\begin{enumerate}
\item \textbf{Data providers.}
Traditionally, geographic data was collected exclusively by experts and professionals in large public and private institutions.
As Web 2.0 and VGI have emerged, a new category of non-expert users/producer (`produsers') has entered the production process \cite{coleman:2009:volunteered}.
Crowdsourced primary sources include contributions from a wide variety of information producers, ranging from experts operating within public and private institutions to non-expert, unpaid, pro-active users.

\item \textbf{Primary sources.}
Projects such as Wikipedia and \gls{osm} collect a large amount of information about the world through crowdsourced efforts.
On the other hand, primary sources such as WordNet are expert-authored, while other projects combine both crowdsourcing and expert control.
Most \glspl{kb} rely heavily on these primary sources, often aligning and merging them into larger knowledge bases.
Inconsistencies and contradictions in primary sources can be propagated onto the derived \glspl{kb}.
For example, an incorrect piece of information in a Wikipedia article will be also found in DBpedia and YAGO.
For this reason, assessing the quality of these primary sources bears particular importance (see Section \ref{sec:quality}).

\item \textbf{\Glspl{gkb}.}
Typically, open \glspl{kb} consist of structured and aggregated versions of existing semi-structured or unstructured primary sources. 
However, some datasets lie at the boundary between primary sources and \glspl{kb}, as they are both interlinked with existing \glspl{kb} and produce new data through crowdsourcing and expert contributions (e.g. Freebase and OpenCyc).
Several \glspl{kb} encode the same primary data into different formalisms, such as DBpedia and YAGO.
\end{enumerate}


These three actors are part of an open system, in which more or less
structured data flows in complex patterns that determine the nature,
quality and limitations of the resulting projects.  Investigations on
such collaborative open processes have been carried out, both in the
area of general crowdsourcing and \gls{vgi}
\cite{coleman:2009:volunteered}.
The next section covers applications
where these \glspl{kb} play an important role, in particular in
relation to ontology alignment, and \gls{gir}.

\section{Open Geo-Knowledge Bases in action}
\label{sec:action}
 
Since the late 1990s, geospatial \glspl{kb} have been of fundamental importance in many geographic applications \cite{winter:2001:ontology}, including 
semantic geographic information systems \cite{abdelmoty:2007:building}, \gls{gir} \cite{odon:2010:geographical}, and toponym disambiguation \cite{overell:2008:using}. 
In general, \glspl{gkb} are used to achieve semantic interoperability between local geographic datasets modelled on incompatible ontologies.  
\Glspl{gkb} can also be useful in cases where advanced geographic
knowledge is necessary to interpret unclear, fuzzy spatial information
queries and needs
\cite{harvey:1999:semantic,giunchiglia:2010:geowordnet}. We focus initially on ontology alignment and merging (Section \ref{sec:aligning}), and we subsequently discuss the usage of ontologies in GIR (Section \ref{sec:gir}).
Our own contributions to this area of research are presented in Section \ref{sec:osmsemnetwork}.

\subsection{Mapping, aligning, and merging Geo-Knowledge Bases}
\label{sec:aligning}

Online geo-data is stored in many different formats, leading to a radical heterogeneity of data formats, ontologies, semantic models \cite{fonseca:2002:using}.
Standards by the Open Geospatial Consortium (OGC) and International Organization for Standardization (ISO) are used in geospatial modelling, while other standards are developed and promoted by the World Wide Web Consortium (W3C) in the context of web technologies.
To date, despite some initial efforts, there is no clear sign of convergence towards broad adoption of joint standards between those two communities \cite{schade:2008:standardizing}.
From an information retrieval perspective, the issue of coverage is critical, as GISs want to access geo-data from as many sources as possible in a consistent way.
For these reasons, the field of integration of \glspl{gkb} has received a lot of attention, and is currently an active research area \cite{lopez:2011:geo,parundekar:2010:aligning}.

The integration of heterogenous data sources relies on the semantic matching of \glspl{gkb}, which present similar conceptualisations of geographic entities using different syntax, structure, and semantics \cite{vaccari:2009:geo}.
According to Giunchiglia et al., considering ontologies as graphs, their alignment consists of the production of a ``set of correspondences between the nodes of the graphs that correspond semantically to each other'' \cite[p. 1]{giunchiglia:2007:semantic}.
Combining \glspl{gkb} poses unresolved challenges, mostly due to
the dynamic nature of online open datasets, semantic ambiguity, and
inconsistent use of the same vocabulary, which is defined by Vaccari et al. as the `semantic heterogeneity problem' \cite{vaccari:2009:geo}.  

Such an integration can operate at the class level (e.g. identify and inter-link the concept `lake' in all the data sources), or at the instance level (e.g. find the entities representing Lake Ontario in all of the ontologies).
Moreover, the semantic mapping can be applied to data, such as geo-ontologies, vector datasets, and gazetteers, or to services, in particular web services, focusing on the public interface signatures \cite{fonseca:2002:using,janowicz:2010:semantic}.
Choi et al., surveying this area of research, have defined three categories of ontology mapping.
The first category includes mapping between local ontologies and a higher-level ontology.
The second category of mapping is performed between local ontologies.
The third category of mapping is part of ontology merging, in which existing ontologies are combined in a bigger ontology \cite{Choi:2006:survey}.

In order to identify semantically close classes and instances in different ontologies, semantic similarity measures are particularly useful.
Semantic similarity measures specific for geographic classes have been surveyed by Schwering \cite{schwering:2008:approaches}.
She classifies the existing measures into geometric, feature, network, alignment, transformational models.
Janowicz et al. have proposed a formal framework for geo-semantic similarity \cite{janowicz:2011:semantics}.
This new framework responds to the ambiguity and lack of clear theoretical grounds that characterise the area of semantic similarity measurement.


Some form of ontology alignment and merging, either partly or fully automatic, has been utilised to generate most of the \glspl{gkb} surveyed in Section \ref{sec:survey}.
The process leading to the creation GeoWordNet, for example, relies on the alignment between GeoNames and WordNet at the class level.
Some of the concepts modelled by GeoNames were not defined in WordNet, prompting the creation of new synsets.
After the ontologies were aligned at the class level, it was possible to align them at the instance level, resulting in the new, integrated ontology \cite{giunchiglia:2010:geowordnet}.

Similarly, LinkedGeoData has mapped some of its instances to corresponding entities in DBpedia, by aligning the ontologies along feature type, spatial distance, and name similarity \cite{Auer:2009:linkedgeodata}.
The ontology YAGO is assembled by aligning WordNet synsets with the less structured Wikipedia articles \cite{Suchanek:2007:yago}.
Along similar lines, Buscaldi et al. have linked existing gazetteers with WordNet, and Wikipedia \cite{buscaldi:2006:inferring}.
Their system extracts place names from the freely available  Geonet Names Server (GNS) and the Geographic Names Information System (GNIS).
Relevant place names are then filtered and enriched using semantic knowledge in Wikipedia and WordNet.
The particular challenge that this system addresses is the combination of semantically flat place names with nodes in complex semantic networks.  


The system GeoMergeP uses a layered architecture to combine and merge local ontologies, through the ISO 19100 standard \cite{buccella:2010:geomergep}.
Surveying recent ontology integration works, Bucella et al. identify three main techniques, which at times are used in isolation, and at times in combination: (1) top-level ontology, (2) logical inferences, and (3) matching/similarity functions.
GeoMergeP combines all of the three approaches to overcome their limitations. 

While most approaches are top-down, the integration can be a bottom-up process.
A bottom-up ontology alignment, focused on geographic linked data, has been carried out by Parundekar et al. \cite{parundekar:2010:aligning}.
This work adopts the approach of common extension comparison, i.e. two classes in different ontologies are considered similar if they are linked by similar instances, to align DBpedia, GeoNames, and LinkedGeoData (see Section \ref{sec:survey}).
The mapping is done through an alignment hypothesis, built bottom-up, starting from instance pairs, up to the most general classes in the ontology.

Ontology alignment is also used by Smart et al. to combine multiple gazetteers through a common, high-level ontology \cite{smart:2010:multi}.
Their Geo-Feature Integration module combines toponyms from \gls{osm}, GeoNames, Wikipedia, and other sources into a unified gazetteer.
The module relies on spatial and textual similarity to match places across the selected data sources.
In addition to traditional text similarity measures, this system uses the SoundEx algorithm to match phonetically similar sounding terms to detect alternative -- and wrong -- place name spellings.



Once the \glspl{gkb} have been integrated, they can be used to support various spatial tasks.
In particular, GIR has emerged as a prominent area that can benefit from \glspl{gkb} \cite{jones:2006:gir}.
The next Section surveys recent work in the area of \glspl{gkb} applied to GIR. 

\subsection{Ontology-powered Geographic Information Retrieval (GIR)}
\label{sec:gir} 

Information retrieval (IR) is a vast and rapidly evolving area of computer science.
Manning et al. define it as ``finding material (usually documents) of an unstructured nature (usually text) that satisfies an information need from within large collections'' \cite[p. 1]{manning:2008:introduction}.
Users' information needs often include some spatial information, such as a geo-location, a street name, and so on.
In the context of ever-growing online information, the geographic dimension of information has become a promising way to increase the chances of meeting information needs.

Traditionally, search engines have treated spatial-related terms as any other textual information.
Over the past decade, however, the area of \gls{gir} has emerged to develop techniques to give geographic information a special treatment, increasing the system's geographic intelligence \cite{jones:2001:geographical}.
Geographic information is often implicit in the documents: broad
geographic entities are omitted when they are assumed to be known to the readers, e.g. Ireland is not mentioned when referring to Dublin in the Irish media.
Toponyms (place names) also have a high degree of semantic ambiguity, as there are many terms to indicate the same geographical entity in different cultural and social groups, and several places have the same name (e.g. more than 40 North-American towns are called Greenville).
Moreover, toponyms pose particular challenges across different natural languages, where historical and spelling variations are very common \cite{Overell:2009:geographic}.

\Glspl{gkb} have been identified as a promising support tool to develop more sophisticated GIR systems \cite{jones:2006:gir}.
While describing their project YAGO, Weikum et al. advocate the usage of knowledge bases to go beyond the limitations of current keyword-based search engines \cite{weikum:2009:database,Suchanek:2007:yago}.
As they put it, the main challenge is ``how to extract the important facts from the Web and organize them into an explicit knowledge base that captures entities and semantic relationships among them'' \cite[p. 61]{weikum:2009:database}.
To illustrate YAGO's knowledge representation, the entity representing Max Planck is displayed, including geographic knowledge about the town where the German physicist was born.
The underlying intuition is that geographic knowledge is generally not provided explicitly, therefore knowledge bases can be used to discover implicit connections between entities.

Ontologies have been used in information retrieval to increase the system intelligence.
In \gls{gir}, Lutz and Klien described an ontology-based system \cite{lutz:2006:ontology}.
A shared vocabulary is used to translate queries across multiple ontologies, without defining a full global ontology.
Their GIR system allows for user-friendly queries, translating generic queries to specific, local \glspl{gkb}.
This is accomplished in a transparent way using Description Logics (DL), a family of knowledge representation languages based on first-order logic that has gained popularity in Semantic Web applications \cite{baader:2005:description}.
Fouad et al. have devised a location-based service to retrieve semantic information based on the user's location \cite{fouad:2010:location}.
Their application performs keyword-based queries on DBpedia, LinkedGeoData, and GeoNames, with the aim of displaying semantically enriched web maps.
Furthermore, in the area of location-based services, DBpedia Mobile demonstrates the possibility of obtaining rich semantic information about the user's surroundings \cite{becker:2008:dbpedia}.

Among others, we regard the following areas to be particularly promising as application domains for \glspl{gkb}:

\begin{description}
\item[\textbf{Named Entity Recognition and Classification (NERC).}]
Several systems rely on NERC techniques to identify location, people and organisations names in raw text.
Nadeau and Sekine surveyed NERC techniques, from the field inception in 1991 to 2006, and discussed the main strategies to evaluate them \cite{nadeau:2007:survey}.
While most NERC approaches are at least partly supervised, Cimiano and V{\"o}lker have developed an ontology-based, unsupervised NERC procedure \cite{cimiano:2005:towards}.
Overell has developed an approach to recognise and classify geo-referenced entities in Wikipedia articles \cite{Overell:2009:geographic}.

\item[\textbf{Toponym disambiguation.}] 
As many geographic locations share the same name, resolving the correct referent in a given context is far from being trivial.
Toponym disambiguation is a specific case of proper name disambiguation, where the proper names refer to explicit or implicit spatial relationships.
Knowledge-based techniques exploit \glspl{gkb} \cite{buscaldi:2011:approaches}.
For example, Overell and R{\"u}ger have utilised Wikipedia as a knowledge base to perform place name disambiguation \cite{overell:2008:using}.
A co-occurrence model is extracted from Wikipedia to provide not only a list of synonyms for each location, but also the context in which each synonym is used.
Toponym disambiguation is tightly connected to the issue of toponym resolution.

\item[\textbf{Toponym resolution.}]
By definition, a toponym refers to a geographic location.
For this purpose geo-coding (toponyms to locations) and reverse geo-coding (locations to toponyms) services have been built using open \glspl{gkb}, extending traditional gazetteers with richer semantic structures.
Smart et al., for example, have integrated several toponym sources into an ontological geo-coding and reverse geo-coding service \cite{smart:2010:multi}.
In order to associate geographic locations to entities, Odon et al. have extracted textual evidence from Wikipedia articles \cite{odon:2010:geographical}.
Using Wikipedia as a semantic network, the importance of the entities is assessed by their connectivity with other entities.
In this way, a representation of the geographic content of Wikipedia articles can be obtained.
A related task is that of `spatial co-reference resolution,' i.e. determining whether two digital representations refer to the same real-world entity.
De Tr\'e et al. developed approaches to detect co-referent features based on possibility theory, and applied it to the issue of duplicate detection \cite{bronselaer:2009:semantical,detre:2010:consistently}.

\item[\textbf{Spatial footprints.}]
Text or multimedia documents can be associated with a spatial footprint, which can be a simple geo-coordinate, a minimum bounding rectangle, or a complex polygon.
Suitable spatial footprints can be computed and indexed, allowing for efficient retrieval and combination with pure text-based indexing. 
Fu et al. have devised an ontology and footprint-based query expansion mechanism \cite{Fu:2005:ontology}.
Spatial entities are identified in a geographic ontology, and the spatial footprint of terms is computed and used in the retrieval process.
Similarly, Vaid et al. have described different indexing approaches for text documents, showing that spatial indexing can enrich pure textual indexing to search large collections of text documents \cite{vaid:2005:spatio}.
In the same area, Martins et al. discuss the `geo-scope' of a text document, which is essentially a spatial footprint to be matched against the query footprint \cite{martins:2005:indexing}.

\item[\textbf{Spatial reasoning.}]
In information retrieval, queries are often expressed in natural language.
Words such as `in' and `near' can convey important spatial semantics, which should be taken into account to meet users' information needs.
GIR systems can utilise ontologies to carry out inferences on such geospatial hints.
For example, the fuzzy query `lakes near Dublin' could be translated into a `within' spatial query with a radius appropriate to the user context.
Fu et al. have conducted work in this area in the frame of the project SPIRIT (Spatially-Aware Information Retrieval on the Internet) \cite{Fu:2005:ontology}.
SPIRIT relies on dedicated ontologies to interpret spatial relationships between query terms, in the format $\langle what, relationship, where\rangle$.
Valid relationships include, among others, `near', `north', and `outside of'.
More recently, the user context emerged as an important element that should be handled by GIR systems. 

\item[\textbf{User context.}]
Any information need, whether containing a geographic dimension or not, is relative to a user context.
For example, a user might want to retrieve all the Italian restaurants in London to conduct a socio-economic analysis, or simply to go out for dinner.
The user context contains diverse information about the user, such as interests, current location, habits, language, computational device, etc, and can be exploited to refine semantic similarity measures and GIR \cite{kessler:2007:similarity}.
Ke{\ss}ler et al. have devised a semantic language to enrich OWL with context sensitivity \cite{kessler:2009:semantic}. 
The \emph{Geo-Finder} system extracts fuzzy spatial footprints from text documents, determining the scope of the search based on the user location and speed \cite{bordogna:2011:geographic}.
The spatial context is taken into account by Mobile Geotumba, a GIR system optimised for handheld devices, to retrieve local information \cite{freitas:2006:mobile}.
\end{description}

All of the aforementioned areas of research are active, and open \glspl{gkb} reviewed in Section \ref{sec:survey} can provide useful tools to explore novel approaches.
Despite the promising results obtained in several works, such ontologies have clear limitations with respect to quality that should not be ignored by researchers and general users alike.
The quality and limitations of \glspl{gkb} will be discussed in Sections \ref{sec:quality} and \ref{sec:limitations}.

The next section describes our contribution to the area of \glspl{gkb}, in particular presenting the \gls{osn} and a semantic enhancement technique to link \gls{osm} entities to DBpedia.


\section{The OSM Semantic Network}
\label{sec:osmsemnetwork}
\glsreset{osm}
In this section we describe our own contribution to the area of \glspl{gkb} and information integration.
The \gls{osn} is a resource that we have extracted from \gls{osm} data to provide a semantic support tool. 
\gls{osm} is the leading project of VGI, and its vector dataset has been discussed, evaluated, and utilised in various contexts \cite{Haklay:2008:openstreetmap,haklay:2010:good}.
The \gls{osn} is extracted through a dedicated web crawler we have developed, and provides a detailed representation of the conceptualisation underlying \gls{osm}.

In \gls{osm}, the semantics of map entities is described through tags, fragments of text with a key and a value (e.g. \emph{amenity=park}, \emph{name=`Central Park'}).
Such tags are proposed, defined, and discussed on a wiki website, which hosts detailed definitions and usage guidelines for the project.\footurl{http://wiki.openstreetmap.org/wiki/Map\_Features}
In the wiki pages, users often link the OSM tags to similar concepts in Wikipedia.
Overall, the tagging process is deliberately informal and open to revision.
Contributors are encouraged to stick to well-known tags, but the creation of new tags is not discouraged, resulting in highly dynamic -- and often inconsistent -- semantics.

In the context of the Linked Open Data, LinkedGeoData (LGD) has converted and published the \gls{osm} vector dataset in RDF, linking it to a formally defined ontology \cite{Auer:2009:linkedgeodata}.
However, the LinkedGeoData ontology is a simple, shallow tree representing tags.
To the best of our knowledge, the rich semantic information on the \gls{osm} wiki website has not been included.
In order to fill this knowledge gap, we have developed an open source tool, the OSM Wiki Crawler, which extracts an RDF graph from the OSM Wiki website.
The crawler extracts a semantic network in RDF, whose vertices represent tags, and edges relationships between tags.
Tags are linked to Wikipedia pages, and to existing LinkedGeoData classes. 
The edge labels specify a number of different relationships between vertices, ranging from a generic internal link (\texttt{link}) to a logical implication (\texttt{implies}).
The detailed content of the current RDF graph is summarised in Table \ref{table:osmwikigraph}.
In addition to the OSM Wiki Crawler, pre-extracted RDF graphs are available online.\footurl{http://wiki.openstreetmap.org/wiki/OSMSemanticNetwork} 
Among other applications, this ontology can be used as a support to compute semantic similarity between tags \cite{janowicz:2011:semantics}, as well as aligning \gls{osm} and LinkedGeoData to other \glspl{gkb} \cite{giunchiglia:2010:geowordnet}.


\begin{table}[t]
\begin{tabular}{ p{18em} p{15em} r }
\noalign{\smallskip}\svhline\noalign{\smallskip}
\emph{RDF Prefix} & \emph{Vertex Type} & \emph{Instances}\\
\noalign{\smallskip}\svhline\noalign{\smallskip}
{\tt osmwiki:Key:} & OSM Key. & 884\\
{\tt osmwiki:Tag:} & OSM Tag. & 2,047\\
{\tt osmwiki:Proposed\_features} &OSM Proposed Tag. & 340 \\
other &LGD and Wikipedia nodes.* & 1,398  \\
\noalign{\smallskip}\svhline\noalign{\smallskip}
\emph{RDF Prefix} & \emph{Edge Type} & \emph{Instances}\\
\noalign{\smallskip}\svhline\noalign{\smallskip}
{\tt osmwiki:link} & Internal link. & 11,982 \\
{\tt osmwiki:valueLabel} & A value of a OSM tag. & 2,926 \\
{\tt osmwiki:keyLabel} & OSM key. & 2,251 \\
{\tt rdf:rdf-schema\#comment} & OSM Tag description. & 1,892 \\
{\tt osmwiki:key} & Link to OSM key page. & 1,891 \\
{\tt osmwiki:combinedWith} & Tag is combined with target tag. & 1,257 \\
{\tt osmwiki:link} & A link to a Wikipedia page. & 1,118 \\
{\tt osmwiki:redirect} &Redirect to a OSM wiki page. & 478 \\
{\tt osmwiki:implies} &Tag implies target tag. & 97 \\
\noalign{\smallskip}\svhline\noalign{\smallskip}
\end{tabular}
\caption{The \gls{osn} (extracted on \graphdate{}). Vertices marked with $^*$ are leaf vertices, i.e. have only incoming edges. `{\tt osmwiki:}' stands for {\tt http://wiki.openstreetmap.org/wiki/}}
\label{table:osmwikigraph}
\end{table}
 
In the context of ontology alignment, we have developed an integration technique between LinkedGeoData and DBpedia, matching geographic features across the datasets.  
As discussed in Section \ref{sec:survey}, some LinkedGeoData instances are linked to corresponding nodes in DBpedia, in particular cities, airports, lakes, and other well-defined entities.
This alignment was performed in the context of pessimistic assumptions, favouring precision over coverage.
As a result, only a small subset of \gls{osm} objects is linked to DBpedia.
Thus, to obtain a wider coverage, we have adopted more flexible heuristics, based on geographic proximity and a tag matching mechanism based on key words.
A web application was built to allow users to visually explore the \gls{osm} dataset, and extract DBpedia nodes and concepts related to the geographic entities displayed in the current web map.
A preliminary evaluation, published in \cite{Ballatore:2011:semantically}, suggests a promising performance of this ontology-based system, but further work is needed to explore its strengths and weaknesses.


 


\section{The quality of crowdsourced Geo-Knowledge Bases}
\label{sec:quality}

In order to utilise open \glspl{kb} successfully, it is crucial to assess their quality with respect to the user's requirements.
For example, a \gls{gkb} might have sufficient quality to enrich the semantics of a web-based \gls{gir} system, but is likely to fail to meet the standards needed by the transport industry.
Assessing the quality of \glspl{kb} can benefit project owners, contributors, users, indicating criteria to select the best available resource for a given task and highlighting limitations and design flaws.

A crucial trade-off in the \glspl{gkb} discussed in this survey is between coverage and precision.
Wikipedia-based ontologies such as DBpedia and YAGO cannot aim at pristine perfection, but can still obtain a reasonable precision \cite{Suchanek:2007:yago}.
On the other hand, expert-authored resources such as WordNet have very high precision, but are unable to compete with the coverage of crowdsourced projects.
A similar trade-off applies to the geospatial dimension: \glspl{gkb} can either reach high, expert-validated spatial quality, or can be updated very frequently by a large number of volunteers, but it is difficult for these two elements to co-exist.

In recent years, several quantitative approaches to assess the quality of an ontology have been discussed \cite{gomez:2001:evaluation,brank:2005:survey,strasunskas:2008:empirical,gangemi:2006:modelling,staab:2004:evaluate,guarino:2004:toward}.
In our view, the approaches to evaluate the quality of \glspl{gkb} can be classified in four families:
 
\begin{enumerate}
  \item \textbf{Manual evaluation}: domain experts and intended users analyse manually the \gls{kb}, highlighting issues and giving qualitative judgements on the mapping between the \gls{kb} and the real world domain that the \gls{kb} is supposed to capture \cite{giles:2005:internet}.
  Although human subjects can easily detect design flaws in the schema, the labour cost of human experts can make it impractical.
  Moreover, even in the presence of considerable resources, large \glspl{kb} cannot be fully evaluated manually, and automatic methods are needed.
  
  \item \textbf{Within-knowledge-base evaluation}: particular properties of a \gls{kb} are observed without comparison with external sources.
  These approaches are based on relationship patterns, distributional patterns, and logical inconsistencies \cite{burton:2005:semiotic,tartir:2005:ontoqa}.
  Although this approach is inexpensive, and can be adopted for any \gls{kb}, its effectiveness is largely context-dependent.
  The average connectivity between objects, for instance, can vary across different domains, without being a reliable indicator of quality.
 
  \item \textbf{Between-knowledge-base evaluation}: two \glspl{kb} covering the same domain can be compared, cross-checking their quality.
  If one of the two \glspl{kb} has comparatively high quality, it can be used as a `ground truth' or `gold standard' to validate the other.
  For example, datasets collected and validate by national mapping agencies tend to obtain higher spatial accuracy than equivalent crowdsourced data \cite{haklay:2010:good}.
  Clearly, this approach cannot be used when comparable \glspl{kb} are unavailable, a rather common situation.
  
  \item \textbf{Application-based evaluation}: ultimately, \glspl{kb} are designed and populated to provide support for real-world applications.
  Hence, an approach consists of observing the performance of a task with and without a given \gls{kb}, and measuring the differential as an indicator of quality.
  In this framework, different \glspl{kb} can be compared indirectly, bearing in mind that \glspl{kb} can obtain varying performances on different tasks.
  Strasunskas and Tomassen have proposed a scheme to evaluate the `ontology fitness' with respect to search tasks \cite{strasunskas:2008:empirical}.     
\end{enumerate}

In practice, quality assessment strategies can combine these four approaches in different ways, along multiple dimensions. 
The quality of a \gls{kb} can be measured at the class level and at the instance level, looking at the statistical properties of the knowledge base.
For example, it is possible to have a solid, well-designed schema but noisy, insufficient instances, or vice-versa.
A combination of these two aspects can offer a comprehensive picture of the ontology quality.

Specific approaches to ontology evaluation focus on a set of dimensions.
Tartir et al. \cite{tartir:2005:ontoqa}, for example, outlined a within-ontology approach based on a triangular model, in which three dimensions of quality can be observed: between the real world and the schema, between the real world and the knowledge base, and between the schema and knowledge base.
In their formulation, metrics for schema quality include `relationship richness,' `attribute richness,' and `inheritance richness,' while instance metrics capture `class importance,' `cohesion,' `connectivity,' and `readability.'   
Logical inconsistencies in the knowledge base can also be detected and used to measure quality \cite{arpinar:2006:ontology}.
For example, a \gls{kb} can contain the conflicting statements `Canada \emph{southOf} USA' and `USA \emph{southOf} Canada.'

Moreover, Burton-Jones et al. addressed the issue of ontology quality from a semiotic viewpoint, proposing a within-ontology evaluation framework \cite{burton:2005:semiotic}.
The quality is observed from four perspectives: `syntactic quality' (richness of lexicon and correctness),  `semantic quality' (interpretability, consistency and clarity), `pragmatic quality' (comprehensiveness, accuracy and relevance), and `social quality' (authority and history).
An overall indicator of quality is obtained with a linear combination of these four dimensions. 

In the context of the open \glspl{gkb} that we have described in Section \ref{sec:survey}, the quality of primary sources such as Wikipedia and \gls{osm} has a great impact of the derived ontologies.
The reliability of Wikipedia has fostered a major academic and intellectual debate, without reaching a monolithic verdict \cite{magnus:2009:trusting}.
A typical way of assessing the quality of Wikipedia is based on a between-knowledge-base comparison of a random sample of articles against a well-established, expert-authored encyclopedia \cite{giles:2005:internet}.
The results indicate that Wikipedia has excellent coverage, but the quality of its articles can vary from poor to excellent.
Hu et al. have proposed within-ontology quality measures for Wikipedia articles, based on the authoritativeness of the contributors \cite{hu:2007:measuring}.

Although false information, hoaxes and spam are generally corrected in a timely manner, Wikipedia articles at a given time can always have errors being introduced and removed.
Therefore, when a snapshot of the Wikipedia website is stored and analysed, any particular article might happen to be captured right after being vandalised or after a thorough revision by a domain expert.
To date, no easy solution to this issue exists.

Assessing the quality of geographic data is a well known area of GIScience, traditionally developed in the framework of cartography \cite{burrough:1996:spatial,hunter:2009:spatial}.
Several dimensions can be observed to assess the quality of spatial information, including positional accuracy (how accurate the object location is with respect to the real world), completeness (how many objects are represented in the map versus all the existing objects), and logical consistency (duplicate objects, inconsistent topological relationships, etc.).
Moreover, semantic aspects are particularly important for GIR, such as attribute and semantic accuracy, which focus on the quality of the metadata. 
The temporal quality, i.e. the rate and accuracy of updates, bears particular importance in several geospatial applications \cite{haklay:2010:good}.

Indeed, the advent of \gls{vgi} introduces additional challenges.
The quality and reliability of \gls{osm} has been debated since its inception, and is now considered a critical research area for VGI \cite{Bishr:2007:geospatial,mooney:2010:towards}.
Like other crowdsourced projects, OSM has experienced recurring and extensive vandalism, urging the project founder to call for action \cite{coast:2010:enough}.
Allegations have been put forward that some vandalism might be carried out by corporate competitors \cite{garling:2012:google}. 
This sort of `spatial vandalism' in open data poses peculiar challenges for project administrators, and has not yet been studied on a systematic basis. 

Analogously to Wikipedia, precision and coverage of the OSM spatial data can vary greatly.
An approach to quantify quality consists of adopting a map from a trusted source (e.g. a national mapping agency), and comparing it with OSM.
Thus, Haklay have compared a sample from the OSM vector dataset against the corresponding data from the British Ordnance Survey \cite{haklay:2010:good}.
OSM obtains a positional accuracy of 70\%, with drops to 20\%, a range that Haklay considers to be ``not dissimilar to commercial datasets'' \cite[p. 700]{haklay:2010:good}.
Along similar lines, Mooney et al. have conducted a quality analysis on a European subset of OSM \cite{mooney:2010:towards}.
Their study confirms the high variability in the data quality, identifying several geographical divides: rural and low-income areas tend to have lower coverage than wealthy, urban areas; natural features tend to be less covered than man-made features.   

To date, the lack of standardised `fitness' metrics to indicate the quality of open \glspl{gkb} makes their adoption problematic, particularly in areas in which the requirements are strict, e.g. logistics and transport.
However, mainly for economic reasons, a number of online services are moving from commercial Web maps to VGI data sources -- the popular social network FourSquare being the most prominent case -- indicating a rising trust in crowdsourced geographic data \cite{hardy:2012:facing}.
 

\section{Current limitations of Geo-Knowledge Bases and GIR}
\label{sec:limitations} 
  
Given the promise of \glspl{gkb}, \gls{gir} and the Semantic Web in general, it is important to be aware of the current limitations and drawbacks of such technologies. 
The Semantic Web is a broad and ambitious project that has made undeniable progress, but many of its issues are largely unresolved \cite{Millard:2010:consuming}.
Polleres et al. have identified critical problems affecting the web of Linked Open Data, ranging from cases where there is too little data, poor quality data, or too much data \cite{Polleres:2010:can}.
We identify issues affecting the usability of linked \glspl{gkb}, restricting the discussion to aspects relevant to the ontologies reviewed in Section \ref{sec:survey}.

\begin{description}
\item[\textbf{Ambiguity.}]
Because of the wide variety of data linked by ontologies, the same vocabulary can have very different usages depending on the context.
A paradigmatic case is the {\tt owl:sameAs} predicate, which has become ambiguous in real datasets \cite{halpin:2010:owl}.
The difficulties in specifying geographic information share a common root in the complexity of the concept of place in natural languages.
The conceptualisation of place is a cultural and language-dependent process, is intrinsically vague, refers to ever-shifting cultural borders, depends on other complex concepts, and is influenced by the context of usage \cite{Santos:2006:place}.
Moreover, the web of open data lacks a meta-ontology framework to describe ontologies in a unified way \cite{jain:2010:linked}.

\item[\textbf{Coverage.}]
In some cases there is too little data, and missing entities or links prevent queries from retrieving results.
When an entity has not been published in RDF and loaded in a public triple repository, it is simply unreachable.
RDF adoption online is sparse, and most RDF triples are coming from mass imports from unstructured or semi-structured datasets \cite{Polleres:2010:can}.
When using open ontologies, the coverage/quality dilemma has to be taken into account: increasing coverage normally entails a drop in quality, and vice-versa.
Projects aiming at global coverage often stumble upon the difficulty of keeping large \glspl{kb} in the same coherent semantic framework.
Coverage also varies depending on fine-grained, project-specific aspects.
In \gls{osm}, for example, man-made features are generally better covered than natural features \cite{mooney:2010:towards}.
The coverage of the interlinking between ontologies, can also show high variability, leaving vast areas of ontologies unlinked \cite{Polleres:2010:can}.

\item[\textbf{Quality.}]
Most \glspl{gkb} contain a vast amount of data imported from crowdsourced projects.
As discussed in Section \ref{sec:quality}, while crowdsourcing has clear advantages in terms of coverage and cost, precision is inevitably neglected.
Moreover, when inconsistent, incomplete or inaccurate information is entered in Wikipedia or \gls{osm}, it will be propagated into DBpedia, YAGO, LinkedGeoData, and many other derived ontologies.
The quality of VGI and crowdsourced data in general is hotly debated, and high variability has to be expected (see Section \ref{sec:quality}).
The difficulties related to creating, maintaining and interpreting metadata were bluntly but persuasively described in the `Metacrap' article by Doctorow \cite{Doctorow:2001:metacrap}.
However, several ontology quality metrics have been devised.
Stransunkas and Tomassen, while presenting a framework for ontology evaluation for information retrieval, survey existing ontology metrics \cite{strasunskas:2008:empirical}.
Beside formal quality metrics based on structure, coherence, and other aspects, an open geo-ontology can be evaluated indirectly on the basis of results obtained in real-world tasks.

\item[\textbf{Expressivity.}]
Modelling geographic knowledge into an ontology poses specific challenges.  
RDF triplification, however simple it might be in most cases, can be very complex and counter-intuitive for certain facts \cite{Polleres:2010:can}.
This issue is due to well-known representational limitations of semantic networks.
Additionally, OWL expressivity for spatial data is very limited.
As Abdelmoty et al. pointed out, OWL does not support spatial types, and common spatial operations such as distance are not available \cite{abdelmoty:2007:building}.
For spatial reasoning, OWL has to be used in conjunction with spatial databases, preventing a seamless integration with existing infrastructures \cite{dolbear:2006:owl}.


\item[\textbf{Complexity.}] Spatial reasoning has been often identified as a fundamental instrument to increase intelligence in GIS \cite{egenhofer:1995:naive}.
However, applying complex spatial reasoning over large \glspl{gkb} poses remarkable challenges.
Even in an ideal situation -- data without noise and logical contradictions -- reasoning in OWL Full is undecidable, and OWL DL is not designed to reason over massive, distributed datasets \cite{Polleres:2010:can}.
Further research is needed to enable efficient spatial reasoning in noisy, large, distributed \glspl{kb}.
\end{description}

Several efforts are being undertaken to tackle these issues in the context of the Linked Data initiatives \cite{Polleres:2010:can}.
However, it is reasonable to assume that the presence of noise, varying quality, and limited expressivity can be reduced but never fully resolved.
Therefore, when developing applications relying on open \glspl{gkb}, caution is needed in order to deal with unexpected contradictions, inconsistencies, ambiguity, and a varying amount of noise in the data.


A prominent application area for \glspl{gkb}, GIR is a relatively young discipline, and its achievements are particularly difficult to assess \cite{Martins:2005:challenges}. 
Most of the works in the area present a preliminary evaluation, leaving the effectiveness of the approaches to be verified empirically in real world applications.
To date, the most important large-scale evaluation is represented by the four GeoCLEF challenges, run from 2005 to 2008 \cite{gey:2006:geoclef,mandl:2008:geoclef}.
Focus on open data was put in GikiCLEF 2009, an evaluation contest conceived to explore cultural and linguistic issues in Wikipedia-based GIR \cite{santos:2010:geographic}.

The driving intuition behind such initiatives is that adding geographic knowledge to an IR system would improve its performance when dealing with information needs with a spatial component.
However, as Mandl noted, complex GIR systems have not consistently obtained better results than geographically naive systems \cite{mandl:2011:evaluating}.
According to Leveling, the contradictory results of GeoCLEF show possible areas of research that might improve the overall results of GIR, strengthening the usage of natural language processing with semantic indexing, handling metonyms, and topological relations beyond simple inclusion \cite{leveling:2011:challenges}.

\section{Conclusions and future work}
\label{sec:concl}

In this chapter we presented a survey on recent advances in open \glspl{gkb} and GIR.
In Section \ref{sec:digearth}, we framed these areas of research in
the combined visions about the Digital Earth, the \gls{sgw},
and the emergence of Volunteered Geographic Information, which have changed the face of geographic information over the past decade \cite{Egenhofer:2002:toward,Goodchild:2007:citizens}.
The linked open geo-resources available online that we discuss in this chapter are realising, at least in part, the vision of the `collaboratory,' a collaborative geo-laboratory envisaged by Al Gore in 1998 to promote the development of geographic digital technologies \cite{gore:1998:digital}.

A survey of free, open \glspl{gkb} with global coverage was presented in Section \ref{sec:survey}, including GeoNames, DBpedia, YAGO, GeoWordNet, ConceptNet, and others.
Those \glspl{kb} are created by extracting knowledge from Wikipedia, \gls{osm} and traditional GIS data sources, merging different \glspl{kb}.
Section \ref{sec:aligning} provides an overview of recent work in the area of geo-ontology alignment and merging.
In order to cope with the growing amount of online geographic information, \gls{gir} has emerged.
Section \ref{sec:gir} surveys recent work in the usage of \glspl{gkb} to increase the geographic intelligence of GIR systems. 
Our own contribution to the area of open \glspl{gkb}, the \gls{osm} Semantic Network and an \gls{osm}/DBpedia alignment approach, is subsequently outlined in Section \ref{sec:osmsemnetwork}.

Section \ref{sec:quality} surveyed the existing strategies to assess the quality of \glspl{gkb}, with particular emphasis on the quality of crowdsourced data sources.  
Despite undeniable advances towards the \gls{sgw} and the increased coverage and quality of open \glspl{gkb}, it is important to recognise its current limitations.
Section \ref{sec:limitations} highlights current issues which researchers using open \glspl{gkb} frequently encounter, identifying the core issues in coverage, quality, expressivity, and complexity of \glspl{gkb}.
Similarly, current GIR systems have not met the expected increase in performance over traditional information retrieval, indicating that geographic intelligence needs refinement to become effective in its applications \cite{mandl:2011:evaluating}.

These issues notwithstanding, promising applications of open \glspl{gkb} are to be found in GIR, ontology alignment, toponym resolution, and related areas. 
In this respect, it can be argued that the most effective way to counter scepticism lies not only in formal, academic evaluations such as GeoCLEF, but in the production and dissemination of usable web applications for Internet users.
For this purpose, more collaboration with the human computer interaction community might help devise appropriate interfaces to interact with open geo-data, exploiting these knowledge bases in convincing ways \cite{mandl:2011:evaluating}.
Work on open \glspl{gkb} should never lose contact with the ultimate stakeholders in information retrieval systems, the human users with their diversified and often unexpected information needs. 

 
\begin{acknowledgement}
The research presented in this paper was funded by a Strategic Research Cluster grant (07/SRC/I1168) by Science Foundation Ireland under the National Development Plan. The authors gratefully acknowledge this support.
\end{acknowledgement}


\bibliography{thesis,mypub}
\bibliographystyle{svmult/styles/spmpsci}

\end{document}